\documentclass[pra,twocolumn,showpacs,preprintnumbers,amsmath,amssymb]{revtex4}

\usepackage{graphicx}% Include figure files
\usepackage{dcolumn}% Align table columns on decimal point
\usepackage{bm}% bold math
\usepackage{color}

\def\vfi{\varphi}
\begin{document}

%\preprint{APS/123-QED}

%\title{Turbulence in nonlocal media: From dispersive shocklets to giant collective incoherent shock waves}% Force line breaks with \\
\title{On the origins of spectral broadening of incoherent waves:\\ Catastrophic process of coherence degradation}
%:\\

%\author{authors}
\author{G. Xu,$^{1,2}$ J. Garnier,$^3$ B. Rumpf,$^4$  A. Fusaro,$^1$ P. Suret,$^2$ S. Randoux,$^2$ A. Kudlinski,$^2$ G. Millot,$^1$ A. Picozzi,$^1$}
\affiliation{
$^1$ Laboratoire Interdisciplinaire Carnot de Bourgogne (ICB), UMR 6303 CNRS -- Universit\'e Bourgogne Franche-Comt\'e, F-21078 Dijon, France\\
$^2$Univ. Lille, CNRS, UMR 8523 - PhLAM - Physique des Lasers Atomes et Mol\'{e}cules, F-59000 Lille, France\\
$^3$Centre de Mathematiques Appliqu\'ees, Ecole Polytechnique, 91128 Palaiseau Cedex, France\\
$^4$Department of Mathematics, Southern Methodist University, Dallas, TX 75275, USA
}


\begin{abstract}
We revisit the mechanisms underlying the process of spectral broadening of incoherent optical waves propagating in nonlinear media on the basis of  nonequilibrium thermodynamic considerations.
A simple analysis reveals that a prerequisite for the existence of a significant spectral broadening of the waves is that the linear part of the energy (Hamiltonian) has different contributions of opposite signs.
It turns out that, at variance with the expected soliton turbulence scenario, an increase of the amount of `disorder' (incoherence) in the system does not require the generation of a coherent soliton structure.
We illustrate the idea by considering the propagation of two wave-components in an optical fiber with opposite dispersion coefficients.
A wave turbulence approach of the problem reveals that the increase of kinetic energy in one component is offset by the negative reduction in the other component, so that the waves exhibit, as a general rule, a virtually unlimited spectral broadening.
More precisely, a self-similar solution of the kinetic equations reveals that the spectra of the incoherent waves tend to relax toward a homogeneous distribution in the wake of a front that propagates in frequency space with a  decelerating velocity.
We discuss this catastrophic process of spectral broadening in the light of different important phenomena, in particular  supercontinuum generation, soliton turbulence, wave condensation, and the runaway motion of mechanical systems composed of positive and negative masses.
\end{abstract}


\maketitle

\section{Introduction}

Nonlinear and nonintegrable waves tend to evolve towards an increasingly disordered state just as  many other common dynamical systems do: They thermalize, that is they irreversibly approach a state of equilibrium associated to the maximally achievable entropy. 
It may appear to be  counterintuitive that the same mechanism, the increase of disorder or entropy, is responsible for a seemingly opposite phenomenon, namely the formation of large-scale coherent structures. 
%It is our everyday experience that disorder has a natural tendency to increase in nature. This empirical fact can be discussed in the framework of non-integrable Hamiltonian systems of random waves.  Considering the nonlinear Schr\"odinger (NLS) equation as a representative model, an increase of disorder is known to be responsible for a counter-intuitive phenomenon of self-organization, which is ruled by the natural thermalization of the system toward the most disordered state that realizes the maximum of `entropy'. Such a self-organization process manifests itself by the spontaneous formation of a large scale coherent structure, which exhibits different forms depending on the nature of the nonlinearity. In the presence of 
One can distinguish two basic types of self-organization in waves 
depending on the nonlinearity and dispersion: 
For the generic nonlinear Schr\"odinger (NLS) equation with a {\it defocusing} nonlinearity, the large scale coherent structure refers to a homogeneous (plane-)wave, and the self-organization process is known as `wave condensation' \cite{dyachenko92,nazarenko05,PRL05,PD09,PRA11b,
proment12,derevyanko12,onorato14,nazarenko}.
This phenomenon is driven by the irreversible process of thermalization toward the Rayleigh-Jeans  equilibrium spectrum, which exhibits a divergence for the fundamental mode \cite{PRL05,PRA11b}, in analogy with Bose-Einstein condensation in dilute quantum gases \cite{stringari,davis01,bec-c,nazarenko}.
On the other hand, in the presence of a {\it focusing} nonlinearity the coherent structure refers to a soliton (solitary wave), and the corresponding self-organization process was originally termed `soliton turbulence' \cite{zakharov88,zakharov89,jordan,rumpf01,rumpf03,ZakhPhysRep01,
nazarenkoPR,nazarenko,rumpfPRL09,PR14,shalva}.
The soliton turbulence scenario starts from a homogeneous initial condition, then the modulational instability leads to the generation of a train of solitons, which eventually coalesce into a single `big soliton'.
% that remains immersed in a sea of thermalized fluctuations.
In this way the non-integrable system irreversibly relaxes toward an equilibrium state, in which a large scale coherent soliton remains immersed in the midst of small-scale random waves \cite{jordan,rumpf01,rumpf03,ZakhPhysRep01,nazarenkoPR,nazarenko,shalva}.

It is important to underline that both self-organization processes of wave condensation and soliton turbulence have in essence a thermodynamic origin, in the sense that {\it an increase of disorder (entropy) requires the formation of a large scale coherent structure}.
In other words, it is thermodynamically advantageous for the system to generate a large scale coherent structure (homogeneous wave or soliton), because this allows the system to increase the amount of disorder in the form of small-scale thermalized fluctuations \cite{zakharov88,zakharov89,jordan,rumpf01,rumpf03,
ZakhPhysRep01,PRL05,PD09,nazarenko,PR14}.
Notice that this type of conservative self-organization process is of fundamental different nature than different forms of condensation effects recently discussed in dissipative cavity systems \cite{conti08,fischer10,fischer14,PRA11c,
derevyanko12,turitsyn12,turitsyn13,carusotto13}.

It is interesting to discuss this phenomenological behavior of `entropy growth' within a different context, namely the process of supercontinuum (SC) generation that undergo optical waves propagating in highly nonlinear materials \cite{dudley06,skryabin10,dudley_book,agrawal}.
SC generation has been the subject of a significant interest, especially since the advent of photonic crystal fibers.
The high nonlinearity of photonic crystal fibers and the ability to readily engineer their dispersion characteristics offer unique opportunities for the experimental study of incoherent optical waves over large nonlinear propagation lengths. 
SC generation is characterized by a dramatic spectral broadening of the optical wave that usually propagates nearby a zero-dispersion wavelength of a photonic crystal fiber. 
The coherence properties of the SC have been widely studied in different propagation regimes \cite{dudley06,dudley_book,agrawal,genty1,genty2}.
From the point of view of nonequilibrium thermodynamics, the spectral broadening of incoherent waves can be interpreted as an increase of the amount of `disorder' (incoherence) carried by the  optical waves \cite{PR14}.
More precisely, it has been shown that spectral broadening through SC generation can be described in certain specific regimes  as a consequence of the natural thermalization of the optical wave toward the `most disordered' Rayleigh-Jeans  equilibrium distribution \cite{OE09,PRA09,PRA13}. 
It is important to underline that such a nonequilibrium kinetic description has been carried out in the `incoherent regime of SC generation'.
In this regime, the optical wave exhibits a highly incoherent dynamics characterized by very rapid fluctuations that prevent coherent nonlinear effects to take place, such as soliton fission and dispersive wave emission \cite{dudley06,skryabin10,dudley_book,agrawal,
hydro_sc,erkintalo12,silvestre,shalva}, or optical wave breaking \cite{finot08,confortiPRA14,silvestre,heidt17}.
Then at variance with conventional SC regimes, the incoherent regime of SC prevents the  generation of robust and persistent coherent soliton states.
This indicates that the process of spectral broadening of incoherent SC generation does not enter the general class of self-organization processes discussed above, since {\it the generation of broadband disordered waves through incoherent SC does not require the generation of a large scale coherent structure}.

In this article we revisit this problem on the basis of simple nonequilibrium thermodynamic considerations inherited from the wave turbulence (WT) theory.
We show that a prerequisite for the existence of a significant spectral broadening of the waves is that the linear part of the energy (Hamiltonian) has different contributions of opposite signs:
%We show that the presence of a zero-dispersion wavelength in the neighbourhood of the optical spectrum appears as a prerequisite for the existence of a significant spectral broadening of incoherent waves:
Contrary to the self-organization processes discussed above through wave condensation and soliton turbulence, here, an increase of disorder (entropy) in the system does not require the generation of a coherent (soliton) structure.
We illustrate this fact by considering the ideal example of two waves that evolve with opposite dispersion coefficients, i.e., the waves propagate in the normal and anomalous dispersion regimes, respectively.
We show that this system exhibits, as a rule, a phenomenon of virtually unlimited spectral broadening of the waves, in which the increase of kinetic energy in one component is exactly compensated by an opposite reduction of energy in the other component.
At variance with the expected soliton turbulence scenario, this `catastrophic' process of spectral broadening occurs unconditionally, even by considering fully coherent initial states of the waves.
The coherence degradation process is described in detail by nonequilibrium self-similar solutions of the wave turbulence kinetic equations. 
The analysis indicates that the spectra tend to evolve toward a homogeneous distribution in the wake of a front that propagates in frequency space with a velocity that decreases algebraically during the propagation.
From a broader perspective, this work  sheds new light on the originating mechanisms of spectral broadening of incoherent waves, and the inhibition of fundamental processes of self-organization such as soliton turbulence and wave condensation.

\section{VNLS model}

We consider a system of two coupled NLS equations, which is known as a generic model for the description of vector phenomena in optics \cite{AgrawKivshar}, plasma \cite{vnls_plasma}, hydrodynamics \cite{vnls_hydro} or Bose-Einstein condensates \cite{vnls_bec}:
\begin{eqnarray}
 i\partial_z u &=& -  \partial_{tt} u +   (|u|^2 + \kappa |v|^2) u,
\label{eq:nls_u} \\
 i\partial_z v &=& - \eta \partial_{tt} v + (|v|^2 + \kappa |u|^2) v,
\label{eq:nls_v}
\end{eqnarray}
where $u(t,z)$ and $v(t,z)$ denote the amplitudes of the interacting waves. 
As usual in optics, the distance $z$ of propagation in the nonlinear medium plays the role of an evolution `time' variable, while $t$ denotes the retarded time in a reference frame moving with the waves. 
For convenience, we have written the vector NLS (VNLS)  equation in dimensionless form, that is to say, we have normalized the problem with respect to the nonlinear length, $L_{nl}=1/(\gamma N)$, and the `healing' time, $\tau_0=\sqrt{\beta_u L_{nl}}$, where $\gamma$ is the nonlinear coefficient, $\beta_{u}$, resp. $\beta_{v}$, the dispersion coefficient of $u$, resp. $v$, and $N$ is the total power of the waves.
In these units, $\kappa$ denotes the ratio between the cross- and self-interaction coefficients and $\eta=\beta_v/\beta_u$. 
In the numerical simulations we consider periodic boundary conditions over the numerical temporal window $[0,T_0]$. 
%with $T$ much larger than any characteristic length scale of the problem.
The dispersion relations of the waves read $k_{u}(\omega) = \omega^2$, $k_{v}(\omega) = \eta \omega^2$.

The VNLS equation conserves three important quantities: the partial powers 
\begin{equation}
N_\vfi=\frac{1}{T_0}\int_0^{T_0} |\vfi|^2 dt,\label{eq:power}
\end{equation}
with $\vfi=u,v$, and the Hamiltonian, ${H}={E}+{U}$, which has a linear contribution ${E} = {E}_u+{E}_v$:
\begin{equation}
{E}_u(z) = \frac{1}{T_0}\int_0^{T_0}  |\partial_t u|^2 dt, \quad {E}_v(z) = \frac{\eta}{T_0} \int_0^{T_0}  |\partial_t v|^2 dt, 
\label{eq:lin_energy}
\end{equation}
and a nonlinear contribution $U=U_u+U_v+U_{int}$, which can be decomposed into the self-interaction energies
\begin{equation}
U_\vfi(z) = \frac{1}{T_0} \int_0^{T_0} \frac{1}{2} |\vfi|^4 dt,
\label{eq:u_spm}
\end{equation}
with $\vfi=u,v$, and the cross-interaction contribution
\begin{equation}
U_{int}(z) =  \frac{\kappa}{T_0} \int_0^{T_0}  |u|^2|v|^2 dt.
\label{eq:u_xpm}
\end{equation}
Of course the total power $N=N_u+N_v$ is preserved as well, and $N=1$ within our normalization.
Equations (\ref{eq:nls_u}-\ref{eq:nls_v}) are integrable for $\eta=\kappa=1$ (or $\eta=\kappa=-1$) \cite{ZakhInteg}.
Because of the infinite number of conserved quantities, integrable systems of random waves exhibit a fundamental different behavior which is  the subject of a growing interest \cite{OE11,integ_prl,costa14,zakharov15,akhmediev15,
akhmediev16,int_pd,integ_sr}.
In the following we consider the nonintegrable case.

\section{Unconstrained spectral broadening}

\subsection{Preliminary qualitative considerations}

Before entering into a detailed study of the VNLS model, it is instructive to discuss at a purely qualitative level the thermalization of waves whose energy is not positive or negative definite -- as it is the case for the linear energy (\ref{eq:lin_energy}) if $\eta < 0$. 
To be concrete, we  illustrate our purpose by referring back to the examples discussed in the introduction, namely wave condensation, soliton turbulence and supercontinuum generation.
In its basic form, wave condensation can be studied with the defocusing 2D (or 3D) NLS equation, so that both linear and nonlinear contributions to the Hamiltonian can be chosen positive definite, e.g., $H=E+U$, with $E>0$ and $U>0$ \cite{nazarenko05,PRL05}.
Recalling that the kinetic contribution to the energy provides a measure of the spectral width of the random waves ($E$ denotes in substance the second order moment of the spectrum), we see that an increase of $E$ during the propagation necessarily requires a reduction of $U$, since their sum $H$ must be conserved.
Because the formation of a homogeneous plane-wave minimizes $U$ (keeping constant the power $N$), it becomes apparent that a spectral broadening of the field (i.e., an increase of `disorder'), {\it requires} the generation of a plane-wave condensate.
We remark that in this thermalization process, the spectral broadening is limited by the positive definite Hamiltonian $H>0$, so that {\it the increase of kinetic energy is constrained by the reduction of nonlinear energy,} $\Delta E \sim |\Delta U|$. 
Note that this reasoning based on wave condensation can easily be transposed to the focusing regime, where it is now the soliton that realizes the minimum of the Hamiltonian: By generating a soliton the system can increase the amount of disorder (kinetic energy) in the form of small scale fluctuations.

The situation is very different in the case of the incoherent regime of SC generation. 
We consider the  case where the spectrum of the optical wave is close to a zero-dispersion wavelength of the (photonic crystal) optical fiber \cite{dudley06,skryabin10,dudley_book,agrawal}. 
%As already commented, here we do not consider supercontinua generated by highly nonlinear effects such as soliton fission \cite{dudley_book,hydro_sc,silvestre} or optical wave breaking \cite{finot08,heidt17}, because such nonlinear  effects are inhibited by the strongly incoherent regime of propagation \cite{PR14}. 
%SC usually occurs efficiently when the spectrum of the optical wave is close to a zero-dispersion wavelength of the photonic crystal fiber \cite{dudley06,skryabin10,dudley_book,agrawal}. 
The NLS model describing light propagation in the fiber then involves higher-order dispersion effects, so that the corresponding linear energy has multiple higher-order contributions $E=E_2+E_3 +...$, where (the modulus of) $E_2$ refers to the usual second-order dispersion contribution denoting the spectral width of the wave.
This simple observation reveals that, owing to the undetermined signs of higher-order dispersion effects, the energy $E$ is no longer bounded (i.e., positive or negative definite).
Accordingly, the incoherent wave can exhibit a dramatic spectral broadening, since large values of $E_2$ can be offset by a correspondingly large and opposite contribution of $E_3$, without any substantial variation of the nonlinear energy, $U \sim $~const.
It turns out that, at variance with wave condensation, {\it here the increase (variation) of kinetic energy is no longer constrained by the reduction of nonlinear energy}, i.e. we can have $|\Delta E| \gg |\Delta U|$.
Note that the situation is different for supercontinua generated solely by highly nonlinear effects such as soliton fission or optical wave breaking, which can be described in first approximation by the integrable NLS equation (the spectrum is not necessarily close to a zero-dispersion wavelength), see e.g. \cite{hydro_sc,silvestre,finot08}.
In this case the linear kinetic energy is limited to the pure second-order contribution $E=E_2$, so that the increase (variation) of kinetic energy is necessarily constrained by the reduction of nonlinear energy, $|\Delta E| \sim |\Delta U|$, as discussed above for wave condensation (or soliton turbulence).

In the following we discuss the effect of unconstrained spectral broadening of incoherent waves with the ideal example provided by the VNLS model (\ref{eq:nls_u}-\ref{eq:nls_v}). 
This model is particularly interesting in that it captures the very nature of the mechanism that underlies spectral broadening, as revealed by the fact that the spectral broadening process is `catastrophic', in the sense that it occurs unconditionally and it is virtually unlimited.

\subsection{VNLS simulations}

We start to consider numerical simulations realized in the so-called weakly nonlinear regime of propagation, i.e., the highly incoherent regime where the strong randomness of the waves makes linear dispersive effects dominant with respect to nonlinear effects, $|E_\varphi| \gg U$, $\varphi=u,v$.
Note that the strongly nonlinear regime will be discussed later in section~\ref{sec:soliton}, in relation with the role of coherent soliton  structures. 
%We recall that in this regime strongly nonlinear processes, such as modulational instabilities or wave breaking (shock waves), are inhibited by the incoherence of the waves \cite{shock_book,PR14}.

\begin{center}
\begin{figure}[t]
\includegraphics[width=8.8cm]{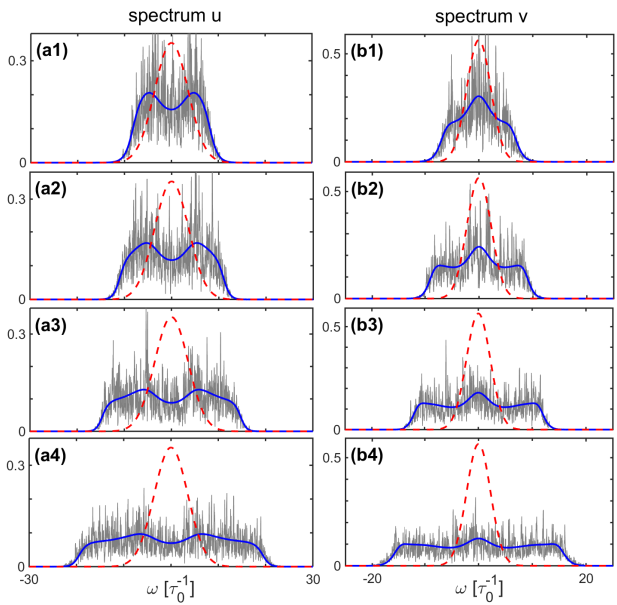}
\caption{Unconstrained spectral broadening: Numerical simulation of the VNLS Eq.(\ref{eq:nls_u}-\ref{eq:nls_v}) (tiny gray), and WT kinetic Eqs.(\ref{eq:kinetic_u2}-\ref{eq:kinetic_v2}) (blue line) showing the evolutions of the spectra of the waves at $z=600$ (a), $z=1600$ (b), $z=4200$ (c), $z=9000$ (d). 
The dashed red line denotes the initial condition.
Parameters are: 
%$\sigma_u \simeq 2.23$ and $\sigma_v \simeq 3.55$, 
$\eta = -1.28$, $\kappa=0.5$.}
\label{fig:evol_nscale}
\end{figure}
\end{center}

\begin{center}
\begin{figure}[t]
\includegraphics[width=8.8cm]{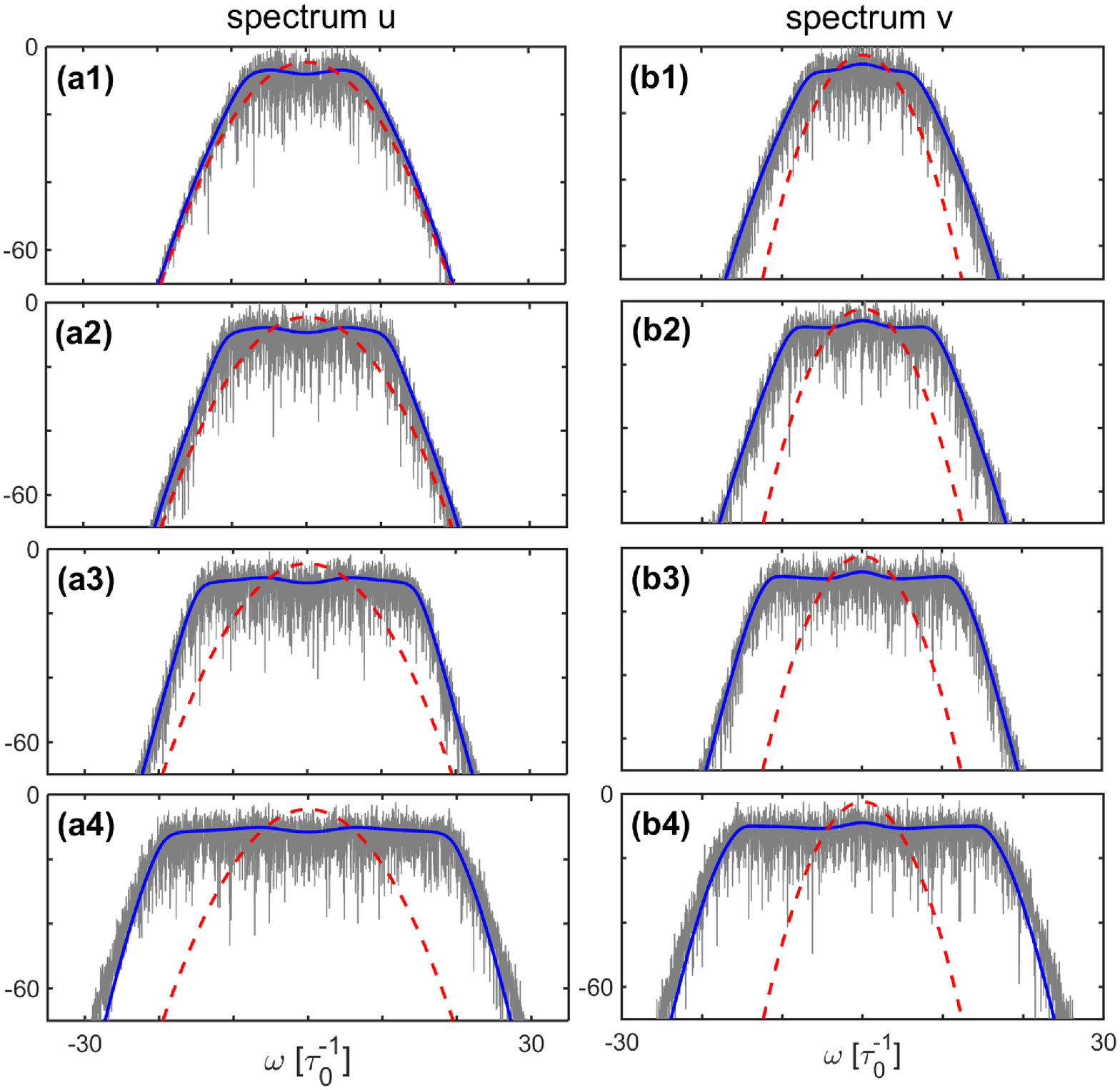}
\caption{Unconstrained spectral broadening: Numerical simulation of the VNLS Eq.(\ref{eq:nls_u}-\ref{eq:nls_v}) (tiny gray), and WT kinetic Eqs.(\ref{eq:kinetic_u2}-\ref{eq:kinetic_v2}) (blue line) showing the evolutions of the spectra of the waves (in 10$\log_{10}-$) scale at $z=600$ (a), $z=1600$ (b), $z=4200$ (c), $z=9000$ (d). 
The dashed red line denotes the initial condition.
The parameters and initial condition are the same as in Fig.~\ref{fig:evol_nscale}: 
%$\sigma_u \simeq 2.23$ and $\sigma_v \simeq 3.55$, 
$\eta = -1.28$, $\kappa=0.5$.}
\label{fig:evol_logscale}
\end{figure}
\end{center}

\begin{center}
\begin{figure}[t]
\includegraphics[width=8.8cm]{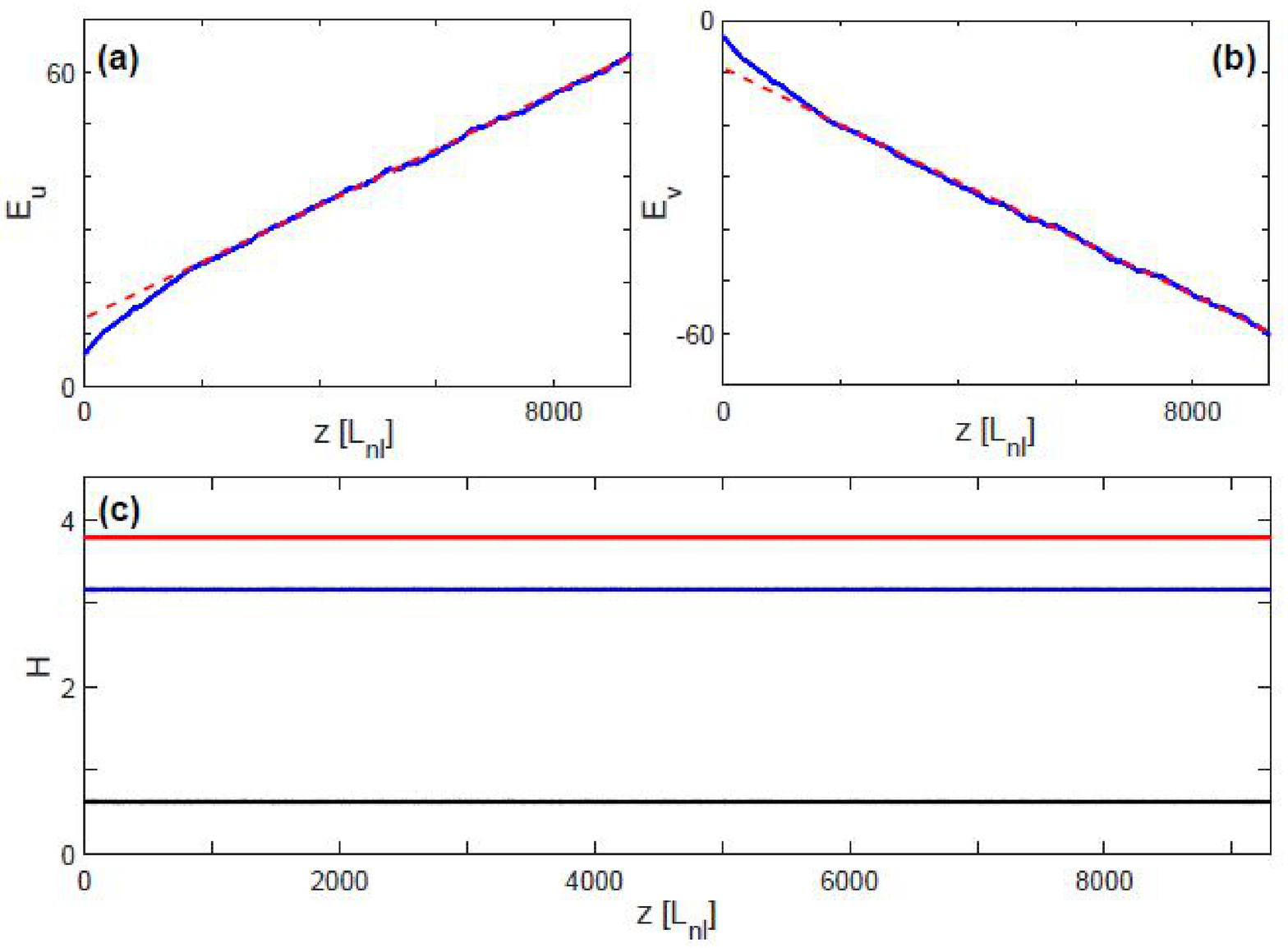}
\caption{Evolutions of the different contributions to the total energy corresponding to the simulation reported in Figs.~\ref{fig:evol_nscale}-\ref{fig:evol_logscale}.
(a-b) Evolutions of the kinetic energies $E_u$ and $E_v$ in blue lines, showing that the energies exhibit linear and opposite evolutions with the propagation length $z$, as predicted by the self-similar behaviour given in Eq.(\ref{eq:kin_en_z}),  $|E_{u,v}| \sim z$ (dashed-red lines).
(c) Evolutions of the sum of the kinetic energies, $E=E_u+E_v$ (blue line),  nonlinear energy $U$ (black line), and total Hamiltonian $H$ (red line).
Note that the variation of the kinetic energies is not constrained by the nonlinear energy, and that $|\Delta E_{u,v}| \gg U$.
Parameters are: $\eta = -1.28$, $\kappa=0.5$.}
\label{fig:ham}
\end{figure}
\end{center}

We assume that the initial incoherent waves have the same power ($N_u=N_v$) and a Gaussian spectrum 
%($\sim \exp[-\omega^2/(2 \sigma_\varphi^2)]$) 
with independent random spectral phases, i.e., they obey a Gaussian statistics and the fluctuations are statistically stationary (homogeneous) in time.
The evolution of the system has already been discussed in the usual case where $\eta > 0$: In the weakly nonlinear regime the two incoherent waves slowly tend to relax toward the expected thermodynamic Rayleigh-Jeans  equilibrium distribution, see e.g. Ref.\cite{PRL10}.
% (note that for $\eta=1$ the system exhibits  a different process of anomalous thermalization \cite{PRL10} that will be discussed below in section~\ref{sec:local_inv}).
In marked contrast with this thermalization process, the simulations with $\eta < 0$ reveal a completely different scenario.
This is illustrated in Figs.~\ref{fig:evol_nscale}-\ref{fig:evol_logscale}, which report the evolutions of the spectra of the waves during the propagation.
The simulations show that both waves exhibit a dramatic spectral broadening, as revealed by the fact that each of the kinetic energies is increased by a factor $\sim 10$.
Clearly, this effect is possible thanks to the opposite dispersion coefficients, so that the increase in the kinetic energy of the $u-$component ($E_u >0$) is compensated by the opposite reduction of kinetic energy in the $v-$component ($E_v < 0$).
% -- note that $|E_\varphi|$ (i.e., the spectral bandwidths) increase typically by a factor $\sim 10$.
More specifically, it is interesting to note that the total kinetic energy is itself almost conserved $E = E_u + E_v \simeq$~const, and then the nonlinear contribution as well, $U \simeq $~const, as reported in Fig.~\ref{fig:ham}.
We note in particular that the variations of the kinetic energies are much larger than the nonlinear energy, $|\Delta E_\vfi|  \gg U$, {\it a feature which is not possible if all energy contributions were positive (or negative) defined.}
It turns out that, at variance with the soliton turbulence scenario, an increase of the kinetic energies does not require the generation of coherent soliton states, as revealed by the analysis of the temporal evolutions of the waves $(u, v)$.
More specifically, we checked that the random waves preserve Gaussian statistics throughout the evolution of the system.
This has been confirmed by considering the evolution of the kurtosis, which is a parameter  that measures the deviation from Gaussian statistics, $K_{\vfi}(z) = \left<|\vfi|^4 \right>/(2 \left<|\vfi|^2 \right>^2) -1$, with $\vfi = u,v$.
The computation of the kurtosis for the simulation reported in Fig.~\ref{fig:evol_nscale}-\ref{fig:evol_logscale} reveals that the variations of $K_{\vfi}(z)$ during the propagation  remain very small, typically smaller than $10^{-3}$.
%${\rm var}(K_{\vfi}) \lesssim 6\times 10^{-4}$.
We have also checked in Fig.~\ref{fig:ham} that, in agreement with Gaussian statistics, we have  $U_\vfi \simeq N_\vfi^2=1/4$ and $U_{int} \simeq \kappa N_u N_v =1/8$. 
Finally note that the role of coherent soliton effects will be discussed later by considering an initial coherent state of the waves, see section~\ref{sec:soliton}.

\section{Wave turbulence approach}

\subsection{Kinetic equations}

The process of dramatic spectral broadening reported above through VNLS simulations can be described in detail by the WT theory, which is the standard mathematical tool to study the dynamics of weakly nonlinear random waves \cite{WT0,ZakhPhysRep01,newellrumpf,nazarenko}. 
This kinetic approach relies on a natural asymptotic closure of the moments equations which is induced by the dispersive properties of the waves.
It leads to a kinetic description of the wave interaction that is formally based on irreversible kinetic equations, a feature which is expressed by a $H$-theorem of entropy growth. 
More precisely, the wave turbulence theory is based on the assumptions that the waves evolve in the weakly nonlinear regime $U/|E_\varphi| \ll 1$, and that they exhibit fluctuations that are statistically stationary in time.
As usual, we also assume that there are no phase-correlations among the random waves $(u,v)$ \cite{ZakhPhysRep01,PRX17}. In this way, the theory derives a set of irreversible kinetic equations that govern the evolutions of the averaged spectra of the fields: 
$\big< {\tilde \vfi}(\omega_1,z) {\tilde \vfi}^*(\omega_2,z) \big> = {n}_\vfi(\omega_1,z) \delta(\omega_1-\omega_2)$, 
%$\big< {\tilde v}(\omega_1,t) {\tilde v}^*(\omega_2,t) \big> = {n}_v(\omega_1,z) \delta(\omega_1-\omega_2)$, 
with ${\tilde \vfi}(\omega,z)=\frac{1}{\sqrt{2\pi}}\int  \vfi(t,z) \exp(-i \omega t) dt$, for $\vfi=u,v$.
%${\tilde v}(\omega,z)=\frac{1}{\sqrt{2\pi}}\int  v(t,z) \exp(-i \omega t) dt$.
Following the wave turbulence procedure, one obtains the following set of coupled kinetic equations \cite{ZakhPhysRep01,PR14}:
\begin{eqnarray}
\partial_z {n}_u(\omega,z) = \frac{\kappa^2}{2\pi} \int \! \! \! \int \! \! \! \int d\omega_1 d\omega_2 d\omega_3 \, {\bf R}_{uv} \, W_{uv}, 
\label{eq:collu}\\
\partial_z {n}_v(\omega,z) = \frac{\kappa^2}{2\pi} \int \! \! \! \int \! \! \! \int  d\omega_1 d\omega_2 d\omega_3 \, {\bf R}_{vu} \, W_{vu},
\label{eq:collv}
\end{eqnarray}
where ${\bf R}_{uv}= {n}_u(\omega_1) \, {n}_v(\omega_2)\,{n}_v(\omega_3)\,{n}_u(\omega)  [ {n}_u^{-1}(\omega)+{n}_v^{-1}(\omega_3)-{n}_v^{-1}(\omega_2)-{n}_u^{-1}(\omega_1) ] $, $W_{uv}=\delta\big(k_u(\omega_1)+k_v(\omega_2)-k_v(\omega_3)-k_u(\omega)\big) \
\delta(\omega_1+\omega_2-\omega_3-\omega)$, while ${\bf R}_{vu}$ and $W_{vu}$ are deduced with the substitutions $u \leftrightarrow v$.
%${\bf N}_{vu}=  n_v(k_1) \, n_u(k_2)\,n_u(k_3)\,n_v(k)  [ n_v^{-1}(k)+n_u^{-1}(k_3)-n_u^{-1}(k_2)-n_v^{-1}(k_1)]$, $W_{vu}=\delta\big(\omega_v(k_1)+\omega_u(k_2)-\omega_u(k_3)-\omega_v(k)\big) \
%\delta(k_1+k_2-k_3-k)$.
%The dispersion relations read $\omega_u(k)=k^2+wk$, $\omega_v(k)=k^2-wk$.
% and $dk_{1-3}=dk_1 dk_2 dk_3$.
The resonant conditions of energy and momentum conservation are expressed by the Dirac $\delta-$functions.
Note that the self-interaction term in the VNLS (\ref{eq:nls_u}-\ref{eq:nls_v}) does not contribute to the kinetic equations, because the conservations of energy and momentum are trivially satisfied in one-dimension.
Equations~(\ref{eq:collu}-\ref{eq:collv}) conserve the powers (`number of particles') of each component, ${N}_\vfi= \int {n}_\vfi(\omega,z) d\omega$, 
%the total momentum, ${\tilde P}=\sum_\vfi {\tilde P}_\vfi$, ${\tilde P}_\vfi=\int k {\tilde n}_\vfi(k,t) dk$ 
and the linear contribution to the Hamiltonian, ${E}= {E}_u+{E}_v$, ${E}_\vfi=\int k_\vfi(\omega) {n}_\vfi(\omega,z) d\omega$ ($\vfi=u,v$). 
The irreversible character of Eqs.(\ref{eq:collu}-\ref{eq:collv}) is expressed by a $H$-theorem of entropy growth, $d{{\cal S}}/dz \geq 0$, where ${{\cal S}}=\sum_\vfi {{\cal S}}_{\vfi}$, and ${{\cal S}}_{\vfi}(z)=\int \log\big({n}_\vfi(\omega,z)\big) d\omega$  is the nonequilibrium entropy of the $\vfi-$th component ($\vfi=u,v$). 
The thermodynamic equilibrium spectra ${n}_\vfi^{RJ}(\omega)$ realizing the maximum of entropy ${{\cal S}}[{n}_u,{n}_v]$, given the constraints of conservation of ${E}$ and ${N}_\vfi$, refer to the well-known Rayleigh-Jeans  distributions
\begin{eqnarray}
{n}_u^{RJ}(\omega)=\frac{{T}}{\omega^2 - {\mu}_u}, \quad 
{n}_v^{RJ}(\omega)=\frac{{T}}{\eta \omega^2 - {\mu}_v}
\label{eq:neqRJ}
\end{eqnarray}
where ${T}$ (`temperature'), and ${\mu}_{u,v}$ (`chemical potentials') are the Lagrangian multipliers associated to the conservation of ${E}, {N}_{u,v}$, respectively.
We recall that in the usual case where $\eta>0$, the tails of the Rayleigh-Jeans  spectra exhibit a power-law behavior $\sim \omega^{-2}$ that reflects the property of energy equipartition among the modes: For large $\omega$ we have $\varepsilon_\vfi(\omega) = k_\vfi(\omega) {n}_\vfi^{RJ}(\omega) \simeq T \ge 0$.

\subsection{Analysis of Rayleigh-Jeans  equilibrium\\ with $\eta < 0$}
\label{subsec:rj}

It proves convenient for the following discussion to introduce Lagrange multipliers associated to the conservation of $E$ and $N_{u,v}$ that are independent of each other, namely $\beta=1/T$ and ${\alpha}_{u,v}=\beta \mu_{u,v}$, so that the Rayleigh-Jeans distributions read ${n}_u^{RJ}(\omega)=1/(\beta\omega^2 - {\alpha}_u)$ and 
${n}_v^{RJ}(\omega)=1/(\beta\eta \omega^2 - {\alpha}_v)$.
Possible values of $\beta$ and ${\alpha}_{u,v}$ follow from the fact that ${n}_{u,v}^{RJ}$ is non-negative for all $\omega$. 
We first remark that ${n}_{u,v}^{RJ}(\omega=0)\ge 0$ requires ${\alpha}_{u,v}< 0$. In addition, the constraints on the parameters that follow $\beta\omega^2 - {\alpha}_u>0$ and $\beta\eta \omega^2 - {\alpha}_v>0$ depend on the sign of $\eta$.

(i) We first discuss the case $\eta>0$, which requires $\beta\ge 0$. 
The waves $u$ and $v$ relax toward the Lorentzian Rayleigh-Jeans  spectrum (\ref{eq:neqRJ}), whose typical spectral bandwidth is given by $\sigma_u \simeq \sqrt{-\mu_u}$, and $\sigma_v \simeq \sqrt{-\mu_v/\eta}$.
We recall that, as discussed in various circumstances \cite{PR14}, the Rayleigh-Jeans  spectrum is not properly defined.
For instance, it is characterized by a divergence of the energy, $E_{u,v} = \int k_{u,v}(\omega) n_{u,v}(\omega,z) d\omega$, a feature which is known to reflect the ultraviolet catastrophe inherent to ensemble of classical waves.
The usual way to regularize such a divergence is to introduce an effective frequency cut-off in the spectrum, i.e., $n_{u,v}^{RJ}(\omega) = 0$ for $|\omega| > \omega_c$. 
Note that such a frequency cut-off $\omega_c$ arises naturally in numerical simulations as a consequence of the temporal discretization ($dt$) of the VNLS Eqs.(\ref{eq:nls_u}-\ref{eq:nls_v}), i.e., $\omega_c = \pi/ dt$.
In this way, the power of the $v-$component reads $N_v=(2 T /(\sqrt{-\eta \mu_v})) {\rm arctan}(\omega_c \sqrt{\eta/(-\mu_v)})$, and a similar expression for $N_u$ with the substitutions $\mu_v \to \mu_u$, and $\eta \to 1$.

(ii) Let us now discuss the Rayleigh-Jeans  equilibrium in the presence of a negative dispersion coefficient, $\eta < 0$. 
First of all, it is important to notice that $n_{v}^{RJ}(\omega) \ge 0$ requires $\beta=0$ for an unbounded spectral domain without frequency cut-off ($\omega_c \to \infty$). 
At variance with the property of energy equipartition discussed above for $\eta >0$, here the equilibrium with $\beta=0$ corresponds to an equipartition of the power $N_{u,v}$, i.e., $n_{u,v}^{RJ}(\omega) =$~const, with a spectral density of power infinitesimal. 
The fact that the Lagrange multiplier vanishes ($\beta=0$) means that the system is not constrained by the conservation of the energy $E$, because it can produce arbitrary positive and negative amounts of  this quantity in the $u-$ and $v-$component, while conserving the total amount of $E$. 
Note that a similar effect of `unconstrained thermalization' was reported in Ref.\cite{PRX17} for weakly dispersive wave systems (with no second-order dispersion effects) -- the common property between \cite{PRX17} and the present work being the fact that the Hamiltonian of the system is unbounded.
The existence of a homogeneous equilibrium spectrum  
leads to an ultraviolet catastrophe in both components $u$ and $v$, which can again be avoided by introducing a cut-off frequency $\omega_c$.

In the presence of a frequency cut-off $\omega_c < \infty$, the parameter $\beta$ in the Rayleigh-Jeans distribution no longer need to vanish exactly.
Indeed, the constraint that the spectrum must be positive ($n_{v}^{RJ}(\omega) \ge 0$ for  $|\omega| \le \omega_c$), imposes that $-\alpha_v \ge |\eta| \beta \omega_c^2$, i.e., $-\mu_v \ge |\eta| \omega_c^2$.
This is corroborated by the expression of the powers.
While $N_u$ keeps the form given above for $\eta >0$, on the other hand $N_v=(2 T /\sqrt{-|\eta| \mu_v}) {\rm arctanh}(\omega_c \sqrt{|\eta|/(-\mu_v)})$, which confirms that a positive $N_v \ge 0$ requires a large negative chemical potential, $-\mu_v \ge |\eta| \omega_c^2$, i.e., a large spectral width of the $v-$component. 
In addition, while $n_u^{RJ}(\omega)$ keeps a Lorentzian spectral shape, $n_v^{RJ}(\omega)$ diverges to infinity as $\omega \to \omega_c$, i.e., most of the power  of the $v-$component accumulates nearby the frequency cut-off $\omega \lesssim \omega_c$.
Since the frequency cut-off $\omega_c$ is chosen in an arbitrary way, this result is not satisfactory from the physical point of view.
Actually, the subsequent analysis will reveal that the system tends to relax toward a state with $-\mu_{u,v} \gg \omega_c^2$ and $\beta \simeq 0$, which is characterized by  an almost homogeneous (flat) spectrum $n_{u,v}(\omega) \simeq $const that exhibits an equipartition of the power.

\subsection{Simulations of the WT kinetic equations}

The previous qualitative discussion indicates that for $\eta < 0$, the incoherent waves do not exhibit the conventional thermalization process to energy equipartition. Actually, the random waves evolve far away from thermal equilibrium during their propagation.
Hence, in the following we resort to the nonequilibrium description of the system provided by the WT kinetic equations.

In order to get further insight into the the kinetic Eqs.(\ref{eq:collu}-\ref{eq:collv}), we remark that, by exploiting the Dirac $\delta-$functions, they can be reduced to the following form:
\begin{eqnarray}
 \partial_z n_u(\omega,z) &=& \frac{\kappa^2}{4 \pi |\eta| }  \int \frac{{\bf N}_u}{|x -\omega|}   dx,
\label{eq:kinetic_u2} \\
 \partial_z n_v(\omega,z) &=& \frac{\kappa^2}{4 \pi} \int \frac{{\bf N}_v}{|x -\omega|}    dx,
\label{eq:kinetic_v2}
\end{eqnarray}
with 
\begin{eqnarray}
{\bf N}_u = n_u(x) n_v(\omega+f-x) n_v(f) n_u(\omega) \quad \quad  \quad \quad  \quad \quad  \quad \quad   \nonumber \\
\quad \quad \quad   \times \ \Big( \frac {1}{n_u(\omega)} + \frac {1}{n_v(f)} - \frac {1}{n_v(\omega+f-x)} - \frac {1}{n_u(x)} \Big), \\
{\bf N}_v = n_v(x) n_u(\omega+g-x) n_u(g) n_v(\omega) \quad \quad  \quad \quad  \quad \quad  \quad \quad   \nonumber \\
\quad \quad \quad  \times \ \Big( \frac {1}{n_v(\omega)} + \frac {1}{n_u(g)} - \frac {1}{n_v(\omega+g-x)} - \frac {1}{n_v(x)} \Big), 
\end{eqnarray}
where the frequencies $f$ and $g$ are expressed in terms of $x$ and $\omega$ as follows:
\begin{eqnarray}
f &=& \frac{1+\eta}{2 \eta} x + \frac{1-\eta}{2 \eta} \omega, 
\label{eq:phasematch_kin1}\\
g &=& \frac{1+\eta}{2} x + \frac{\eta-1}{2} \omega.
\label{eq:phasematch_kin2}
\end{eqnarray}

The kinetic Eqs.(\ref{eq:kinetic_u2}-\ref{eq:phasematch_kin2}) have been solved by numerical integration, starting from the same initial spectrum considered in the VNLS simulations.
The results are reported in Figs.\ref{fig:evol_nscale}-\ref{fig:evol_logscale} with blue lines.
A remarkable quantitative agreement has been obtained with the simulations of the VNLS (\ref{eq:nls_u}-\ref{eq:nls_v}), without using adjustable parameters.
Note that, in spite of the noisy structure of the VNLS spectra, we can see that they exhibit some global deformations that are reproduced in detail by the simulations of the WT kinetic equations.
%We note that such a good agreement has been obtained even in the tails of the spectrum, down to 10$^{-8}$, as revealed by the plots in logarithmic scale reported in Fig.~\ref{fig:evol_logscale}.
This shows that the phenomenon of unconstrained spectral broadening can be accurately described by the nonequilibrium kinetic formulation provided by the WT theory.

\begin{center}
\begin{figure}[t]
\includegraphics[height=7cm,width=8.5cm]{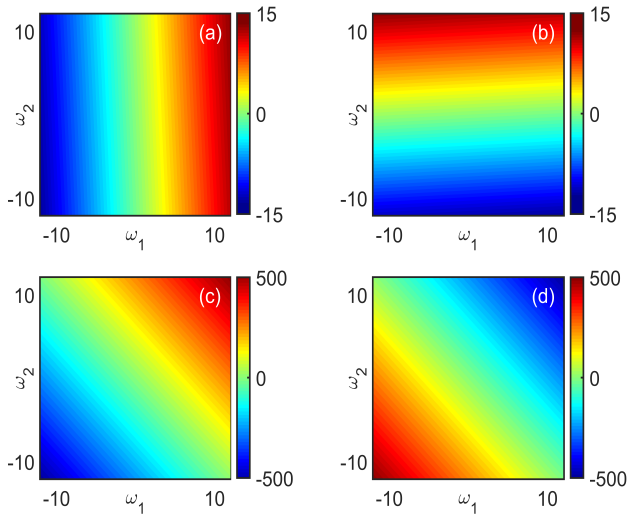}
\caption{Resonant phase-matching conditions of energy and momentum conservation: $\omega_1+\omega_2=\omega_3+\omega_4$, $ k_u(\omega_1)+ k_v(\omega_2) = k_u(\omega_3)+ k_v(\omega_4)$. 
Plots of $\omega_3$ (left column), and $\omega_4$ (right column), as a function of $(\omega_1,\omega_2)$ for $\eta=1.1$ (a-b), $\eta=-1.1$ (c-d), see Eq.(\ref{eq:w3}-\ref{eq:w4}).
For opposite dispersion coefficients, $\eta < 0$, resonant phase-matching is achieved for very large frequency components -- see the different color-bar scales.
For instance, for $\omega_1 \sim \omega_2 \sim 10$, we have the solution $\omega_3 \sim \omega_4 \sim \omega_{1,2}$ for $\eta > 0$ (a-b); while for $\eta < 0$ we have the solution $\omega_3 \sim - \omega_4 \sim 500 \gg \omega_{1,2}$.
}
\label{fig:resonance}
\end{figure}
\end{center}

\begin{center}
\begin{figure}[t]
\includegraphics[width=9.0cm]{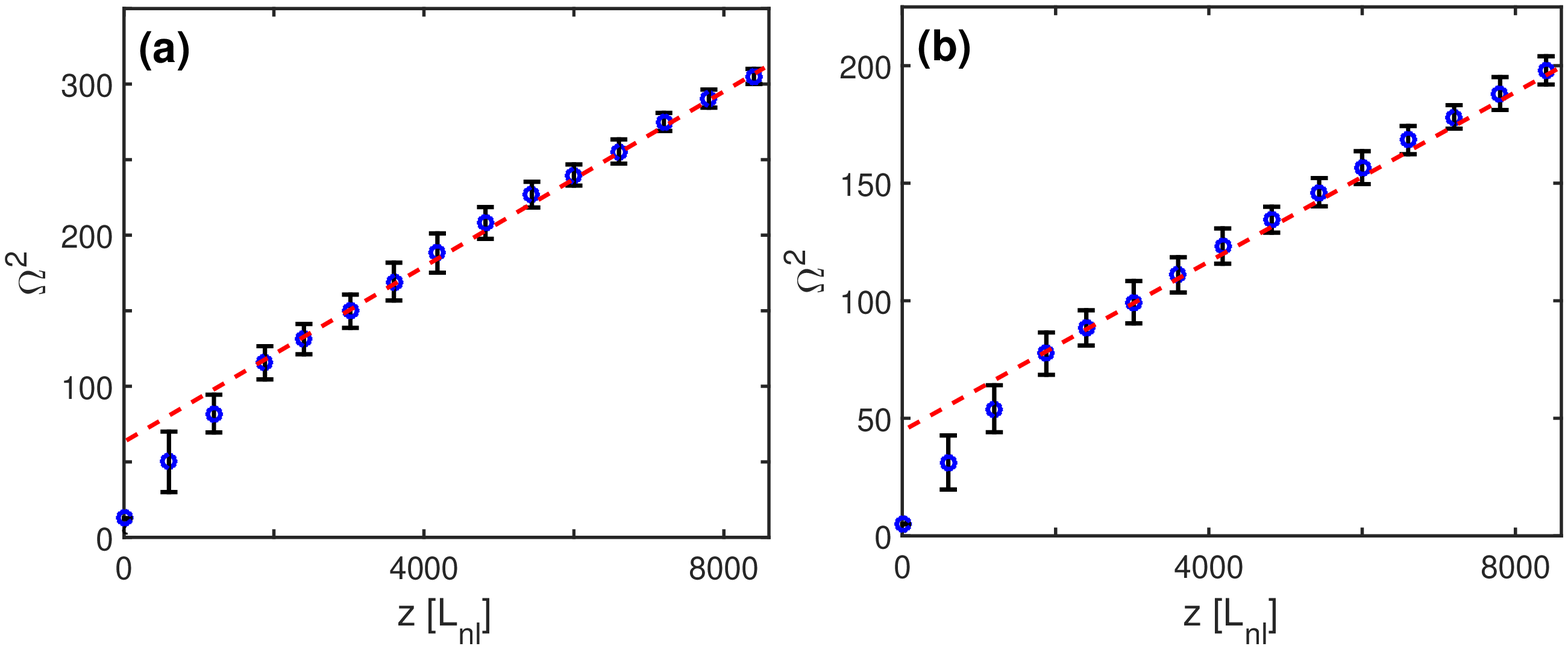}
\caption{Self-similar process of spectral broadening. 
Evolutions of the squares of the spectral bandwidths, $\Omega^2(z)$, for the waves $u$ (a), and $v$ (b) obtained by simulations of the VNLS (the error bars denote the uncertainty in the computation of the spectral widths), and the WT kinetic equations (circles). The linear dependence vs. $z$ (dashed red line) confirms the behaviour predicted from the self-similar solution of the WT kinetic equations $\Omega(z) \sim \sqrt{z}$, see Eq.(\ref{eq:Omega_sqrtz}).
The parameters and initial condition are the same as in Fig.~\ref{fig:evol_nscale}: 
%$\sigma_u \simeq 2.23$ and $\sigma_v \simeq 3.55$, 
$\eta = -1.28$, $\kappa=0.5$.}
\label{fig:Omega_sqrtz}
\end{figure}
\end{center}

\subsection{Phase-matching resonances}
\label{sec:phasematch}

It is instructive to analyze the resonant phase-matching conditions underlying the reduction of the kinetic equations to the form given in Eqs.(\ref{eq:kinetic_u2}-\ref{eq:phasematch_kin2}).
The resonant conditions of energy and momentum conservation read: $\omega_1+\omega_2=\omega_3+\omega_4$, $ k_u(\omega_1)+ k_v(\omega_2) = k_u(\omega_3)+ k_v(\omega_4)$. 
For a given set of two frequencies $(\omega_1, \omega_2)$, their solutions read
\begin{eqnarray}
\omega_3 &=& \frac{1}{1+\eta} \big( (\eta-1) \omega_1 + 2 \eta \omega_2 \big), 
\label{eq:w3}\\
\omega_4 &=& \frac{1}{1+\eta} \big( (1-\eta) \omega_2 + 2 \omega_1 \big),
\label{eq:w4}
\end{eqnarray}
for $\eta \neq -1$ (the degenerate case $|\eta|=1$ will be discussed later).
As  illustrated by Fig.~\ref{fig:resonance}, for $\eta <0$, resonances can be satisfied for very large frequency components, $\omega_{3,4} \gg \omega_{1,2}$, a feature which contrasts with the usual case $\eta > 0$ where the generated frequencies are of the same order as the incident ones, $\omega_{3,4} \sim \omega_{1,2}$.
Note that this can easily be foreseen from an immediate inspection of the analytical expressions (\ref{eq:w3}-\ref{eq:w4}) when $\eta \to -1$.
This simple observation anticipates the possibility of a dramatic spectral broadening of the waves with opposite dispersion coefficients.

\begin{center}
\begin{figure}[t]
\includegraphics[width=9.0cm]{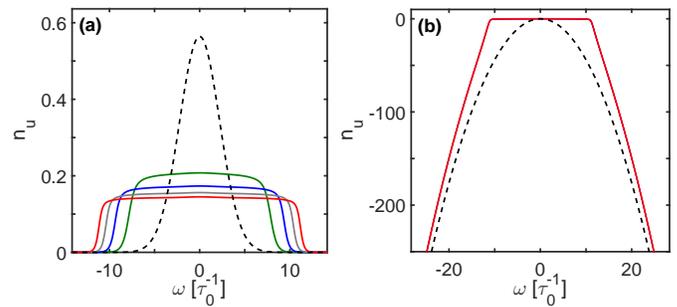}
\caption{Nonlocal nonlinear diffusion in frequency space:
Evolution of the spectrum $n_u(\omega,z)$ obtained by integrating numerically the WT kinetic equations for $\eta=-1$: (a) in normal scale $z=0$ (dashed dark), $z=1000$ (green), $z=2000$ (blue), $z=3000$ (gray), $z=4000$ (red); (b) in $10 \log_{10}-$scale $z=0$ (dashed dark), $z=4000$ (red).
As described by Eqs.(\ref{eq:ke_beta_1_0}-\ref{eq:ke_beta_1}), the spectrum experiences a nonlocal nonlinear diffusion on its initial support and tends to relax toward a homogeneous distribution (i.e., $n_u(\omega) \sim $const on the support of the initial spectrum).
}
\label{fig:nonloc_diff}
\end{figure}
\end{center}

\subsection{Local invariants and nonlocal nonlinear diffusion}
\label{sec:local_inv}

We  comment here on the existence of local invariants revealed by the reduced form of the KE Eqs.(\ref{eq:kinetic_u2}-\ref{eq:phasematch_kin2}).
It is important to remind that such invariants are `local' in the sense that they are verified for each frequency $\omega$ individually.
In this respect, they differ from integral invariants, whose existence leads to a generalized form of the Rayleigh-Jeans  spectrum \cite{PRL10}.

Local invariants were identified in relation with degenerate resonances in Refs.\cite{PRL10,OL10,PR14}.
For instance, for $|\eta|=1$ the phase-matching conditions discussed above exhibit the degenerate solution $(\omega_3=\omega_1, \omega_4=\omega_2)$. 
%More specifically, the reduced form of the KE admit local invariants in the particular case where $|\eta| = 1$.
More specifically, for $|\eta|=1$, Eqs.(\ref{eq:kinetic_u2}-\ref{eq:phasematch_kin2}) admit the local invariant:
\begin{eqnarray}
J(\omega) = n_u(\omega,z) + {\rm sign}(\eta) \, n_v(\omega,z),
\end{eqnarray}
with $\partial_z J(\omega) = 0$ for any frequency $\omega$.
In the case $\eta = 1$, the sum of the two spectra is conserved. Clearly, this constitutes a  severe constraint that inhibits a significant spectral broadening of the waves.

In the opposite case $\eta = -1$, the difference between the two spectra is conserved, so that the waves may exhibit in principle a process of unconstrained spectral broadening, since the broadening of $u$ and $v$ can compensate each other  so as to preserve the invariant $J(\omega)$.
However, the system does not exhibit the unconstrained  spectral broadening discussed above through Figs.~\ref{fig:evol_nscale}-\ref{fig:evol_logscale}. This can be anticipated by the degenerate form of the resonances discussed here above, or by a more detailed analysis of the kinetic equations. Indeed, if the initial conditions are such that $n_v(\omega,0) =  n_u(\omega,0)$, then $n_u(\omega,z)$ satisfies
\begin{eqnarray}
 \partial_z n_u(\omega,z) = \frac{\kappa^2}{2 \pi}
  \Big[ \int \frac{n_u(x)-n_u(\omega)}{|x -\omega|}   n_u(x) dx \Big] n_u(\omega),
\label{eq:ke_beta_1_0}
\end{eqnarray}
where we have assumed that the spectra are even functions. This equation shows that there is no significant spectral broadening, because $n_u(\omega,z)$ appears as a factor of the right-hand side.
In addition, Eq.(\ref{eq:ke_beta_1_0}) reveals an important effect: 
The term between the square brackets describes a nonlinear and nonlocal diffusion of the spectra.
This becomes apparent by writing this term in the form
\begin{eqnarray}
 \int \frac{n_u(x)-n_u(\omega)}{|x -\omega|}   n_u(x) dx = 
 \nonumber \quad \quad \quad \quad \quad \quad \quad \quad \quad \quad \\
 \frac{1}{2} \int \frac{n_u(\omega+y)+n_u(\omega-y)-2 n_u(\omega)}{|y|}  dy n_u(\omega) \nonumber \\
 +
\int \frac{(n_u(\omega+y)- n_u(\omega))^2}{|y|}  dy ,
\label{eq:ke_beta_1}
\end{eqnarray}
where the first dominant term denotes a second derivative in the spatial frequency, while the second term denotes the square of a first derivative, which also describes a broadening of the spectrum.
Therefore, {\it if the initial conditions   exhibit a compactly supported spectral shape, then the spectrum effectively diffuses and tends to become uniform (flat) within this compact domain.}
This effect of nonlocal nonlinear diffusion in frequency space is illustrated in Fig.~\ref{fig:nonloc_diff}, which shows that the broadening of the spectrum $n_u(\omega,z)$ solely occurs on its initial support -- there is no generation of new frequency components as described by the degenerate form of the resonant conditions, $\omega_{3,4}=\omega_{1,2}$ for $\eta=-1$.

This general behavior can be extended to the case where the initial spectra are not identical. In this case, the evolution of $n_v$ is deduced from $n_u$ by $n_v(\omega,z) =  n_u(\omega,z) - J(\omega)$, while $n_u(\omega,z)$ satisfies
\begin{eqnarray}
 \partial_z n_u(\omega,z) &=& \frac{\kappa^2}{4 \pi}
 \Big[ \int \frac{n_u(x)-n_u(\omega)}{|x-\omega|}  \big(2 n_u(x) - J(x) \big) dx \Big] n_u(\omega) \nonumber \\
&& +\frac{\kappa^2}{4 \pi}
 \Big[ \int \frac{n_u(x)-n_u(\omega)}{|x-\omega|}  \big( J(x) - n_u(x) \big) dx \Big] J(\omega)
\nonumber   \\
&& - \frac{\kappa^2}{4 \pi}
\Big[ \int \frac{J(x)-J(\omega)}{|x-\omega|} n_u(x) dx \Big] n_u(\omega).
\label{eq:ke_J_beta_1}
\end{eqnarray}
As in the previous case, this form of the kinetic equation indicates that spectral broadening can  only occur on the support of the initial spectrum. Moreover, we can notice that, while the last term keeps a memory of the initial condition, the first two terms on the right-hand side of (\ref{eq:ke_J_beta_1}) are of the same form as (\ref{eq:ke_beta_1}) and thus also describe a nonlinear and nonlocal diffusion of the spectra.
Accordingly, the spectra $n_\varphi(\omega)$ tend to become homogeneous during their evolutions outside the initial support of $J(\omega)$.

\begin{center}
\begin{figure}[t]
\includegraphics[height=8.5cm,width=9cm]{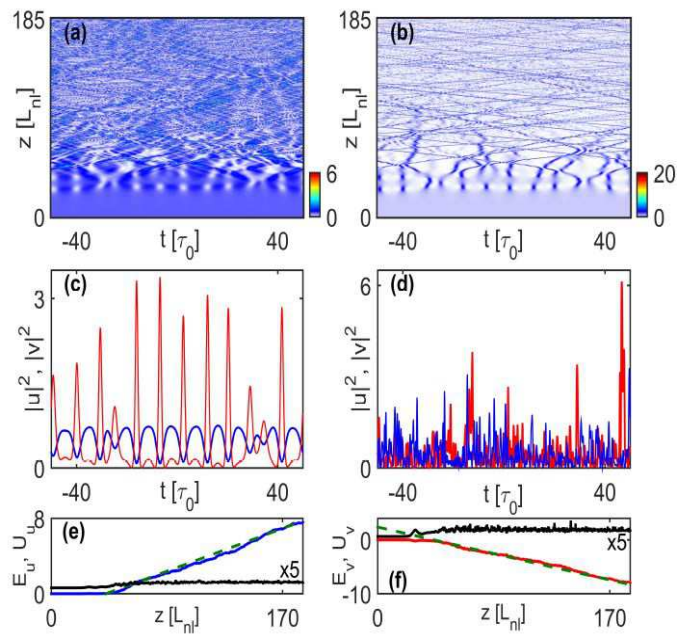}
\caption{Numerical simulation of VNLS Eq.(\ref{eq:nls_u}-\ref{eq:nls_v}) starting from homogeneous (continuous) waves for $u$ and $v$. 
Spatio-temporal evolutions of $|u|^2$ (a), and $|v|^2$ (b), and corresponding temporal profiles at $z=28$ (c), and $z=185$ (d) ($u$ blue line, $v$ red line).
The waves exhibit a modulational instability that leads to the generation of a train of dark-bright solitons for the $u-v$ components, see panel (c).
Evolutions of the linear energies $E_{u}$ in blue line (e), and $E_v$ in red line (f), while the straight dashed green line indicates a linear behavior.
(e-f) Corresponding nonlinear energies $U_{u}$ and $U_v$ in black lines  (multiplied by a factor $\times 5$ for visibility):
The variation of the kinetic energies is not constrained by the nonlinear energies, $|\Delta E_{\vfi}| \gg |\Delta U_{\vfi}|$ for $\vfi=u,v$.
At variance with the expected soliton turbulence scenario, the system eventually enters the highly incoherent regime of interaction characterized by an unconstrained spectral broadening, as indicated by the linear behaviour $E_{u} \simeq - E_v \sim z$ predicted by the theory in Eq.(\ref{eq:kin_en_z}).
Parameters: $\eta=-0.77$, $\kappa=2/3$.}
\label{fig:soliton_turb}
\end{figure}
\end{center}

\section{Self-similar kinetic solutions}

Our aim in this section is to describe some general properties of the process of unconstrained spectral broadening discussed above through Figs.~\ref{fig:evol_nscale}-\ref{fig:ham}.
%We have shown above that for $\eta < 0$ the system does not relax toward the conventional stationary equilibrium RJ distribution.
We have shown that this process occurs very far from equilibrium.
Accordingly, we now consider nonequilibrium self-similar solutions of the kinetic Eqs.(\ref{eq:kinetic_u2}-\ref{eq:phasematch_kin2}).
Because of the conservation of the powers $N_u$ and $N_v$, we look for a self-similar solution in the form
\begin{eqnarray}
n_\varphi(\omega,z) = \frac{1}{\Omega(z)} \Psi_\varphi\Big(\frac{\omega}{\Omega(z)}\Big),
\end{eqnarray}
where $ \Psi_\varphi$ is a normalized profile satisfying $\int \Psi_\varphi(s) ds =N_\varphi$ for $\varphi=u,v$.
By substitution into the kinetic equations we find that the spectral width of the waves $\Omega(z)$ and the spectral profiles $\Psi_u,\Psi_v$ should satisfy
\begin{eqnarray}
\Omega(z) \partial_z \Omega(z) &=& K_0 , 
\label{eq:Omega} \\
K_0 \big( \Psi_u(s) + s\Psi_u'(s) \big) &=& - \frac{\kappa^2}{4\pi |\eta|} \int 
\frac{\bf \Theta_u}{|x' - s|} dx',
\label{eq:Psi_u} \\
K_0\big( \Psi_v(s) + s\Psi_v'(s) \big) &=& -\frac{\kappa^2}{4\pi} \int 
\frac{\bf \Theta_v}{|x' - s|} dx',
\label{eq:Psi_v} 
\end{eqnarray}
with 
\begin{eqnarray}
{\bf \Theta}_u =  \Psi_u(x') \Psi_v(s+f'-x') \Psi_v(f') \Psi_u(s) 
\quad \quad  \quad \quad  \quad \quad  \quad \quad   \nonumber \\
\quad \quad \quad   \times \ \Big( \frac{1}{\Psi_u(x') }+\frac{1}{\Psi_v(f')}-\frac{1}{\Psi_v(s+f'-x')}-\frac{1}{\Psi_u(s)} \Big), \nonumber \\
{\bf \Theta}_v = \Psi_v(x') \Psi_u(s+g'-x') \Psi_u(g') \Psi_v(s) 
\quad \quad  \quad \quad  \quad \quad  \quad \quad   \nonumber \\
\quad \quad \quad  \times \ \Big( \frac{1}{\Psi_v(x') }+\frac{1}{\Psi_u(g')}-\frac{1}{\Psi_u(s+g'-x')}-\frac{1}{\Psi_v(s)} \Big), \nonumber
\end{eqnarray}
for some constant $K_0$, and
%\begin{eqnarray}
$f' = \frac{1+\eta}{2\eta} x'+\frac{1-\eta}{2\eta}s$, 
$g' = \frac{1+\eta}{2} x'+\frac{\eta-1}{2}s$.
%\end{eqnarray}
It is interesting to note that Eq.(\ref{eq:Omega}) provides the self-similar behavior of the process of spectral broadening of the waves:
\begin{eqnarray}
\Omega(z) \sim \sqrt{z}.
\label{eq:Omega_sqrtz}
\end{eqnarray}
This square-root power-law behavior has been confirmed by the numerical simulations of the KE as well as VNLS equations, as reported in Fig.~\ref{fig:Omega_sqrtz}.
An immediate consequence of (\ref{eq:Omega_sqrtz}) is that the kinetic energies of the two waves  grow linearly with the propagation length $z$:
\begin{eqnarray}
|E_\varphi| \sim \int \omega^2 n_\varphi(\omega,z) d\omega =   \Omega^2(z) \int s^2 \Psi_\varphi(s) ds  \sim  z,
\label{eq:kin_en_z}
\end{eqnarray}
for $\varphi=u,v$.
More precisely, Eqs.~(\ref{eq:Psi_u}-\ref{eq:Psi_v}) impose $\int s^2 \Psi_u(s) ds = - \eta \int s^2 \Psi_v(s) ds$, that is the total kinetic energy is conserved $\int \omega^2 n_u(\omega,z) d\omega +\eta \int \omega^2 n_v(\omega,z) d\omega = \mbox{ const}$.
This linear and opposite power-law behavior for the evolutions of the energies $E_u \sim -E_v \sim z$,  has been also confirmed by the numerical simulations, as illustrated in Fig.~\ref{fig:ham}(a-b).

We finally remark that the analysis carried out in the case $\eta=-1$ through Eqs.(\ref{eq:ke_beta_1}-\ref{eq:ke_J_beta_1}) predicts that the spectra exhibit strong nonlinear and nonlocal diffusions. 
In a similar way, for $\eta$ close to $-1$, the spectra $n_\varphi(\omega,z)$ tend to become homogeneous during the propagation, as a result of this process of effective diffusion in frequency space. 
Such a tendency appears to be consistent with the numerical simulations of the VNLS equation and KE reported in Fig.~\ref{fig:evol_nscale}, where we can notice that the spectra tend to become flat in the wake of the front of the spectrum that propagates with the velocity $V(z) = \partial_z \Omega(z) \sim 1/\sqrt{z}$.
Note that this conclusion is consistent with the previous qualitative analysis of the Rayleigh-Jeans spectrum, see section~\ref{subsec:rj}.

\section{Unconstrained spectral broadening from an initial coherent  state}
\label{sec:soliton}

In the previous analysis we have studied the evolution of the system starting from an initial incoherent state characterized by a strong randomness of the waves, i.e., the weakly nonlinear regime $|E_\varphi| \gg U$ at $z=0$ for $\varphi=u,v$.
In this way we have shown that the WT theory properly describes the system throughout its whole evolution.
However, referring back to the soliton turbulence scenario, if the initial condition consists of a coherent state, then one would expect that in its long term evolution the system would self-organize into a large scale coherent soliton that remains immersed in a sea of small scale fluctuations \cite{jordan,rumpf01,ZakhPhysRep01,PLA10}.
Our aim in this section is to see whether the phenomenon of unconstrained spectral broadening can occur even by considering a fully coherent state of the initial waves.
We note that, by starting from a fully coherent state, the system evolves in the strong nonlinear regime of interaction ($U \gtrsim |E_\vfi|$), which cannot be described by the WT theory. Actually, no systematic methods have been developed to describe a strongly nonlinear regime of random waves, whose study is the subject of a growing recent interest in different contexts, see e.g.,  \cite{ZakhPhysRep01,nazarenkoPR,derevyanko12,
nazarenko,rumpfPRL09,PR14,turitsyn13,Lvov_acoustic,
rumpfPRL12,NC15,Silberberg09,PRL04,Josserand13,
vorticesBEC,rumpf15,delre16}.

\begin{center}
\begin{figure}[t]
\includegraphics[height=9.5cm,width=9.5cm]{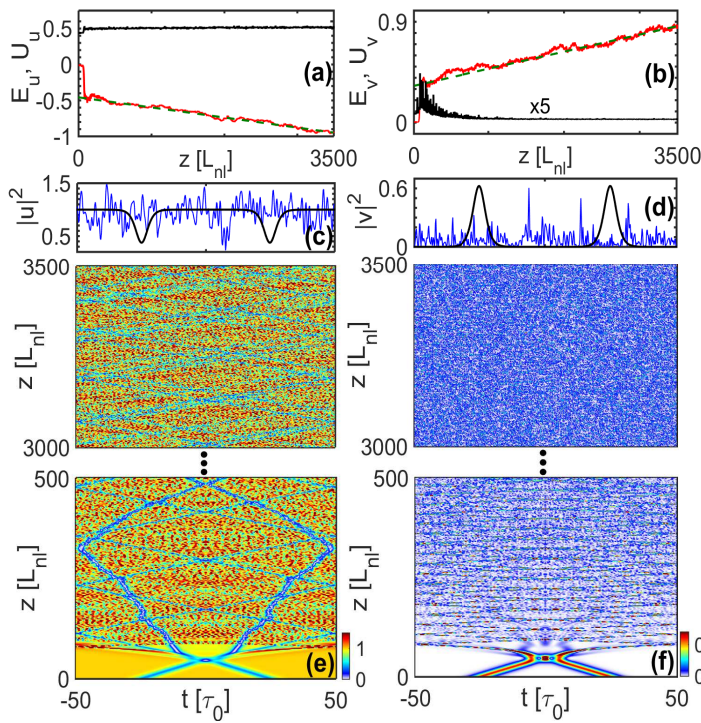}
\caption{Numerical simulation of VNLS Eq.(\ref{eq:nls_u}-\ref{eq:nls_v}) starting from a pair of dark-bright vector solitons for the $u-v$ components at $z=0$.
(a-b) Evolutions during the propagation of the linear energies $E_u$ (a), and $E_v$ (b), in red lines: They follow a linear behavior (dashed green line) as predicted by the kinetic theory, see Eq.(\ref{eq:kin_en_z}).
The dark lines report the corresponding evolutions of the nonlinear energies $U_u$ (a), and $U_v$ (b):
The variation of the kinetic energies is not constrained by the nonlinear energies, $|\Delta E_{\vfi}| \gg |\Delta U_{\vfi}|$ for $\vfi=u,v$.
(c-d) Intensity profiles $|u|^2(t,z=3500)$ (c), $|v|^2(t,z=3500)$ (d) in blue lines, and corresponding initial conditions (dark lines).
(e-f) Spatio-temporal evolutions of $|u|^2(t,z)$ (e), $|v|^2(t,z)$ (f): Despite the weak interaction between the waves imposed by the initial solitonic condition ($N_v/N_u \simeq 0.084$), in the long-term evolution the system tends to evolve toward the strongly incoherent regime that inhibits the generation of robust soliton structures. Parameters: $\eta=-0.77$, $\kappa=2/3$.}
\label{fig:soliton_break}
\end{figure}
\end{center}

\subsection{Homogeneous initial condition}

In  analogy with the soliton turbulence scenario, we have first realized numerical simulations of the coupled VNLS Eq.(\ref{eq:nls_u}-\ref{eq:nls_v}) by considering an initial state composed of coherent homogeneous (continuous) waves, $u(t,z=0)=v(t,z=0)=1/\sqrt{2}$ plus a small random noise (the constant $1/\sqrt{2}$ is due to our normalization, $N_u+N_v=1$).
As illustrated in Fig.~\ref{fig:soliton_turb}, the system exhibits a modulational instability, whose nonlinear stage is characterized by a reversible transfer of power back to the homogeneous waves, thus leading to the well-known phenomenon of Fermi-Pasta-Ulam recurrence mediated by modulational instability   \cite{kuznetsov,kruskal,akhmediev86,fpu_review,onorato_fpu,
haelterman,mussotPR14,erkintalo11,
wabnitz_pla14,kibler15}.
We can observe in particular the formation of a train of soliton-like structures for the coupled $(u,v)$ components.
Since $\eta < 0$, such solitons are of a bright-dark form, as indicated by the intensity profiles of the waves at $z=28$ in Fig.~\ref{fig:soliton_turb}, see panel (c).
Specifically, the vector soliton consists of a bright (sech-type) $v-$component propagating in the anomalous dispersion regime, and a dark (tanh-type) $u-$component propagating in the normal dispersion regime \cite{Kivshar_PLA}. 
Note that, contrary to the `inverted' vector soliton solutions (bright-dark solitons respectively in the normal-anomalous dispersion regimes) \cite{trillo88,lisak90,kivshar91}, here the continuous wave background of the dark component is modulationally stable, since it evolves in the normal dispersion regime. 

We can notice in Fig.~\ref{fig:soliton_turb}(a-b) that the different solitons start to strongly interact with each other.
Although their evolutions are rather robust during the propagation, the non-integrability of the system leads to a complex energy exchange among the solitons and the surrounding fluctuations. 
Solitons can collide, occasionally merge or deteriorate via random interactions, and each of these processes leads to the emission of energy and power to the incoherent wave component. 

At this stage, the evolution of the system is apparently similar to that expected from the general scenario of soliton turbulence \cite{zakharov88,zakharov89,jordan,rumpf01,rumpf03,ZakhPhysRep01,
nazarenkoPR,nazarenko}.
However, owing to the negative dispersion coefficients ($\eta < 0$), in the subsequent evolution the system enters the process of unconstrained spectral broadening, in which the kinetic energies of the waves can increase freely, as confirmed by the linear and opposite evolutions of $|E_{u,v}| \sim z$ in the simulations, see Figs.~\ref{fig:soliton_turb}(e) and (f).
In these figures we also reported the nonlinear self-interaction energies, which tend to approach the corresponding values expected for Gaussian statistics $U_\vfi^G = N_\vfi^2 =1/4$. Note that the deviation from Gaussianity is larger for the $v$ component that evolves in the anomalous dispersion (`focusing') regime, since we have for $z$ large  $U_v \simeq 0.35$, while $U_u \simeq 0.25$.

As a consequence of the free variation of the kinetic energies, the system evolves from the initial strongly nonlinear regime ($|E_\vfi| \ll U_\vfi$ for $z \lesssim 30$), toward the weakly nonlinear regime ($|E_\vfi| \gg U_\vfi$ for $z>100$).
In particular, it is apparent in Fig.~\ref{fig:soliton_turb}(e-f) that the variation of the kinetic energies is not constrained by the nonlinear energy, $|\Delta E_\vfi| \gg |\Delta U_\vfi|$.
In this way, the spectra of the waves continuously spread and the corresponding time correlations decrease during the propagation (see Figs.~\ref{fig:soliton_turb}(d)), so that there is some point at which the highly incoherent nature of the waves can no longer sustain a robust coherent soliton state.   
In other words, even by starting from an initial coherent state, the waves $u$ and $v$ slowly enter a highly incoherent (weakly nonlinear) regime, which eventually inhibits the generation of persistent and robust soliton states.

\subsection{Solitonic initial condition}

To complete our study we discuss the process of unconstrained spectral broadening by considering a   solitonic initial condition.
More precisely, we have considered as initial condition a pair of vector bright-dark solitons, whose analytical solutions were derived in Ref.\cite{Kivshar_PLA}, in particular for $\kappa < 1$.
As commented above, such solitons are of bright (sech-type) and dark (tanh-type) form for the $v$ and $u$ components, respectively (see Fig.~\ref{fig:soliton_break}(c-d) in black lines).
Exploiting the Galilean  invariance of the VNLS equation, we have imposed a velocity to the two solitons, so as to induce a collision between them.
We have also added a small noise (of magnitude $10^{-8}$ for the amplitudes) in the initial condition, which does not qualitatively alter the  dynamics of the waves.
The results of the numerical simulation are reported in Fig.~\ref{fig:soliton_break}.
In the first stage ($z \lesssim 25$), the two solitons propagate against each other with a constant velocity and in an apparent stable fashion, see Fig.~\ref{fig:soliton_break}(e-f).  
The solitons then exhibit an inelastic collision featured by the emission of some radiation, as expected for non-integrable systems. 
In this way, the system slowly enters the incoherent erratic regime discussed  above, which is characterized by a monotonous and unconstrained spectral broadening of the waves.
%Note that, as a consequence of the peculiar resonant phase-matching conditions discussed above through Fig.~\ref{fig:resonance}, the simulations of the VNLS equation with $\eta <0$ are quite sensitive, especially when the random waves evolve in the nonlinear regime as in Fig.~\ref{fig:soliton_break}, so that numerical noise is expected to play some role in the evolution of the system.
We remark in Fig.~\ref{fig:soliton_break}(e-f) that a trace of the soliton trajectories can still be identified in the spatio-temporal evolution, even for relatively large propagation lengths.
This is due to the fact that the interaction between the two components $u$ and $v$ is not efficient in the case considered in Fig.~\ref{fig:soliton_break}, because of the unbalanced amount of powers in the two wave components ($N_u / N_v \simeq 0.084$) that is imposed by the initial solitonic condition at $z=0$.
As a consequence, the nonlinear self-interaction energies are also strongly unbalanced, $U_u \simeq 0.5$ and $U_v \simeq 0.006$ (notice that the corresponding values for Gaussian statistics are $U_u^G = N_u^2 \simeq 0.85$ and $U_v^G = N_v^2=0.006$).
However, despite the fact that $U_v \ll U_u$, we remark in Fig.~\ref{fig:soliton_break}(a-b) that the kinetic energies compensate each other $E_u \simeq -E_v$ , and tend to follow a linear behavior with the propagation length $z$, as predicted by the kinetic wave theory for the process of unconstrained spectral broadening.

\begin{center}
\begin{figure}[t]
\includegraphics[width=9 cm]{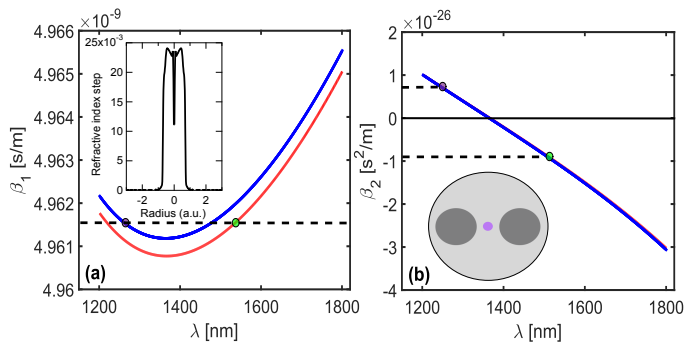}
\caption{Design of the dispersion properties of an optical fiber: Inverse of group-velocity $\beta_1$ vs. wavelength $\lambda$ (a), dispersion curve $\beta_2$ vs. wavelength $\lambda$ (b), for the orthogonal polarization components (blue and red lines).
Waves with $\lambda_1 \simeq 1.27 \mu$m (slow axis, blue curve) and $\lambda_2 \simeq 1.55 \mu$m (fast axis, red curve) propagate with the same group-velocity and opposite group-velocity dispersions, $\eta \simeq -1.2$.
Insets: fully realistic refractive index profile of the fiber core (a); schematic cross section of the birefringent fiber with the core in purple and constraints rod in dark grey (b).
}
\label{fig:exp}
\end{figure}
\end{center}

\section{Discussion and conclusion}

In summary, a simple analysis revealed that a wave system characterized by an unbounded linear energy is expected to exhibit a significant spectral broadening of the waves.
We have illustrated this effect by considering the representative  example of the VNLS model with opposite dispersion coefficients.
We have shown that, as a general rule, the waves exhibit a virtually unlimited process of spectral broadening, in which the kinetic energy in one component can freely increase, since such an increase is balanced by an opposite reduction of energy in the other component, i.e., the increase of kinetic energy is no longer constrained by the reduction of nonlinear energy $|\Delta E_\vfi| \gg |\Delta U_\vfi|$ for $\vfi=u,v$.
As a remarkable consequence, an increase of `disorder' in the system does not require the generation of a coherent structure: 
At variance with the expected soliton turbulence scenario, the  process of coherence degradation occurs unconditionally, even  by considering an initial fully coherent state of the waves.
This catastrophic process of spectral broadening is described in detail by nonequilibrium self-similar solutions of the wave turbulence kinetic equations. 
The analysis indicates that the spectra tend to evolve toward a homogeneous distribution ($n_{u,v}(\omega) \simeq$~const) in the wake of a front that propagates in frequency space with a velocity that decreases algebraically during the propagation ($\sim 1/\sqrt{z}$).
This conclusion is corroborated by the analysis of the Rayleigh-Jeans equilibrium distribution, which reveals that, in principle,  for $\eta < 0$ the system is not constrained by the conservation of the energy $E$, so that the vanishing Lagrange multiplier ($\beta=0$) leads to a constant equilibrium spectrum characterized by an equipartition of power (`particles'), instead of the expected energy equipartition. 
Note that such a homogeneous equilibrium spectrum is not, strictly speaking, a physical state. Its physical importance is its role as a statistical attractor that governs the relaxation process at physically relevant scales.
%It is interesting to note that a similar effect of `unconstrained thermalization' was reported in Ref.\cite{PRX17} for weakly dispersive wave systems (with no second-order dispersion effects) -- the common property between \cite{PRX17} and the present work being the fact that the Hamiltonian of the system is unbounded (not positive or negative definite).

In order to motivate the realization of an experiment, we briefly comment its feasibility by using a highly birefringent optical fiber, in which the waves $u$ and $v$ refer to the linear orthogonal polarization components of the light beam. 
The effect of unconstrained spectral broadening imposes severe constraints on the dispersion properties of the waves. 
However, the dispersion properties and group-velocities can easily be tuned by exploiting the dispersion induced by the fiber waveguide.
We report in Fig.~\ref{fig:exp} a possible realistic design of an optical fiber that exhibits the required dispersion characteristics.
We consider a fiber with a germanium doped core of about 4.5 $\mu$m in diameter (see the refractive index profile in the inset of Fig.~\ref{fig:exp}(a)) surrounded by two boron-doped stress rods (16 mol. \%). The whole fiber cross section in shown in Fig.~\ref{fig:exp}(b). The corresponding first- and second-order dispersion characteristics are reported in panels (a) and (b), respectively.
We note in particular that a wavelength $\lambda_1 \simeq 1.27 \mu$m (generated e.g. by a bismuth-doped fiber laser) and $\lambda_2 \simeq 1.55 \mu$m (generated e.g. by an erbium-doped fiber laser), propagate with the same group-velocity (no walk-off) and disperse with opposite second-order dispersion coefficients, with the ratio $\eta \simeq -1.2$.
Also note that the dispersion curve reported in Fig.~\ref{fig:exp}(b) is relatively flat, which allows one to neglect third-order dispersion effects in a first approximation.
Accordingly, the first stage of  propagation of the orthogonal polarization components in the fiber can be modelled by the VNLS Eq.(\ref{eq:nls_u}-\ref{eq:nls_v}).

It is interesting to note that a somehow similar type of optical fiber system was proposed in Ref.\cite{peschelPRL13} to study diametrically driven self-accelerating pulses propagating at opposite group-velocity dispersions. 
This effect of self-accelerating pulses can be interpreted as an optical realization of a classical diametric drive, where a continuously propulsive effect is achieved by a combination of two fields having effective masses of opposite signs ($\eta < 0$).
Indeed, considering a pair of particles with opposite masses, the gravity of the positive mass will attract the negative mass, while the negative one is repelled.
If properly arranged, the two particles can exhibit an intriguing self-propulsion effect, also see \cite{peschelNP13,longhi14,delrePRL16}. 
In this respect, a statistical mechanics formulation of a system of  particles with both positive and negative masses constitutes by itself a delicate problem.
The general idea is that ensembles of positive and negative masses cannot coexist in a stable equilibrium state because of the underlying self-propulsion runaway motion  \cite{pollard95}. 
In a loose sense, some properties of such a ``catastrophic world" may appear similar to those of the catastrophic spectral broadening of random waves discussed in this work, an analogy that will be the subject of further studies.

\section{Acknowledgements}
A.P. and A.F. acknowledge support from the Labex ACTION (ANR-11-LABX-01-01) program. 
B.R. acknowledges a grant from the Simons Foundation (\# 430192).
A.P. acknowledges funding from the European Research Council under the European Community's Seventh Framework Programme (FP7/20072013 Grant Agreement No. 306633, PETAL project). G.X. acknowledges the support from the ANR project NoAWE (ANR-14-ACHN-0014).
This work has been partially supported by the Agence Nationale de la Recherche through the LABEX CEMPI (ANR-11-LABX-0007) and the Equipex Flux (ANR-11-EQPX-0017), as well as by the Ministry of Higher Education and Research, the Hauts-de-France Regional Council and European Regional Development Fund (ERDF) through the Contrat de Projets Etat-Region (CPER Photonics for Society P4S).

\end{document}